**Origin of the voltage gap and recombination losses in all-perovskite tandem solar cells**

*Hurriyet Yuce-Cakir[1,2*], John F. Roller[3], Haoran Chen[4], Tingting Zhu[4], Susanna M. Thon[2,5,6], Yanfa Yan[4], Zhaoning Song[4], Behrang H. Hamadani[3*]*

[1]PREP Associate, Engineering Laboratory, National Institute of Standards and Technology, Gaithersburg, Maryland 20899, United States.

[2]Ralph S O'Connor Sustainable Energy Institute, Johns Hopkins University, Baltimore, Maryland 21218, United States.

[3]Engineering Laboratory, National Institute of Standards and Technology, Gaithersburg, Maryland 20899, United States.

[4]Department of Physics and Astronomy and Wright Center for Photovoltaics Innovation and Commercialization, The University of Toledo, Toledo, Ohio 43606, United States.

[5]Department of Electrical and Computer Engineering, Johns Hopkins University, Baltimore, Maryland 21218, United States.

[6]Department of Materials Science and Engineering, Johns Hopkins University, Baltimore, Maryland 21218, United States.

Corresponding Author: hurriyet.yucecakir@nist.gov, behrang.hamadani@nist.gov

**Abstract**

All-perovskite tandem solar cells with narrow and wide bandgap perovskite absorbers are promising candidates for low-cost and high efficiency photovoltaic applications. However, the open circuit voltage of typical tandem structures is generally smaller than the sum of the individual voltages in the single-junction form; a quantity we call the voltage gap. Subcell optimization can only begin once nonradiative losses associated with each absorber layer can be properly identified. To address this, we used absolute electroluminescence hyperspectral imaging to construct external radiative efficiency maps of each subcell within the tandem stack and compare these measurements with single junction devices. These measurements were then combined with additional electro-optical characterization and modeling to construct subcell current vs voltage curves. We find that the narrow band gap subcell contributes the most towards the voltage gap and therefore fabrication and processing efforts should focus on reducing nonradiative recombination losses within the narrow band gap absorber.

**1. Introduction**

Over the last several years, single junction perovskite solar cells have emerged as a promising photovoltaic (PV) technology due to the high absorption coefficient,[1,2] low exciton energy,[3] long carrier diffusion lengths,[4,5] tunable band gap energy,[6] and a rapid increase in the performance (power conversion efficiency (PCE) of more than 26.9 %) of halide perovskites.[7] The PCE of single junction perovskite solar cells (PSCs) is approaching the Shockley-Queisser (S-Q) limit of approximately 33 %, and has now prompted researchers to focus on perovskite tandem solar cells that can surpass a *PCE* of 40 %.[6,8]

In all-perovskite tandem structures, multiple perovskite absorber layers with narrow band gap (NBG) and wide band gap (WBG) materials are used to maximize sunlight energy conversion. Mixed tin-lead iodide and mixed halide lead perovskites, having narrow and wide band gaps, respectively, are key components for achieving high efficiencies in tandem perovskite solar cells. However, in order to achieve optimal device performance, good interfacial properties between the subcells are required, and electron-hole nonradiative recombination losses should be minimized. In particular, a key issue that has been identified is related to a *voltage gap* that is observed between the tandem cell's open circuit voltage, $V_{oc}$, and the sum of the individual open circuit voltages of the constituent cells when prepared in single junction (SJ) form, or $V_{oc}^{gap} = V_{oc}^{tandem} - \sum V_{oc}^{SJ}$. A $V_{oc}^{gap} \approx 0$ V indicates that no extra nonradiative recombination losses have been introduced in the tandem structure from those losses that already existed in the SJ forms of each type of absorber material. However, this number is usually a positive number; a voltage gap of ≈ 60 mV has been observed in our earlier tandem cells[9], and only recently, we have shown that in champion tandem devices of the highest quality, it is possible to bring the voltage gap down to ≈ 0 V.[10] For most typical tandem cells where a voltage gap does exist, in order to definitively say whether a given subcell or both equally are responsible for the observed voltage gap, a comprehensive understanding of the physics of charge carrier generation and recombination within each subcell is required.

External radiative efficiency (ERE) has been shown to be an important metric in understanding the extent of charge carrier recombination losses in solar cells.[11–13] The ERE is identified as the fraction of total dark-current recombination that generates radiative emission, evaluated at the $V_{oc}$, and is closely related to the external quantum efficiency at the corresponding injection level.[14] In fact, the so-called $V_{oc}$ deficit, which is the difference between the radiative-limit and the observed open-circuit voltage, can be accurately approximated using ERE measurements.[15] In the literature, ERE is sometimes measured and reported from absolute photoluminescence (PL) measurements, which we refer to as PL-ERE, and sometimes from absolute electroluminescence (EL) measurements or EL-ERE. The PL-ERE is typically conducted on an absorber layer, with or without the full device stack, providing insights into the quasi-Fermi level splitting energy, i.e., the internal voltage of the material.[16,17] Characterization of a whole device, however, requires calculating the ERE across the entire device rather than the absorber layer alone. Therefore, in a full device stack, current can be injected into the device and the luminescence signal from one or more active layers (if the device is a tandem cell structure) can be measured. By measuring the EL-ERE of each subcell simultaneously and establishing the relationship between the ERE and the injection current density, the $V_{oc}$ and injection-dependent losses can be determined at current densities close to the real operational conditions of the cell.[18,19] Additionally, although hyperspectral imaging has been employed in the literature to investigate the optoelectronic properties of single-junction perovskite solar cells,[20,21] comparable studies on perovskite tandem solar cells remain scarce, particularly for examining the individual subcells within a tandem architecture.

In this work, we performed thorough quantitative hyperspectral electroluminescence imaging experiments on NBG and WBG dual-junction all-perovskite tandem solar cells, as well as single junction versions of these same materials. EL-ERE maps of the tandem subcells were carefully compared against single junction ERE maps to quantify the differences between the two, both in terms of the magnitude and also the spatial variations. Subcell ERE maps are found to be significantly more inhomogeneous than their single junction counterparts as quantified by the standard deviation of the pixel magnitudes. Also, the mean ERE values in subcells are at least an order magnitude lower than in single junctions. Separately, extensive dark and light *J-V* measurements were performed on the same devices. To analyze and interpret the measurements, we applied a 2-diode recombination current model, comprised of both radiative and nonradiative components, to our ERE and *J-V* data simultaneously and show that individual subcell *J-V* curves in a two-terminal tandem configuration can be accurately predicted from these combined measurements. Armed with the individual subcell *J-V* curves, we compared the subcell $V_{oc}$ with SJ $V_{oc}$ values and unequivocally show that the bulk of the contribution to the observed $V_{oc}^{gap}$ comes from the extra recombination losses in the NBG subcells.

## 2. Results

### 2.1. Electro-optical characteristics of devices

The device architectures of all-perovskite tandem subcells and the NBG and WBG single junction solar cells (labeled, T-NBG, T-WBG, SJ-NBG, and SJ-WBG) are illustrated in **Figure 1**a and **Figure 1**b, respectively. We measured a statistically significant number of solar cells on several substrates each containing up to 8 functioning devices. Absolute hyperspectral EL photon flux images of several cells on one substrate with seven working cells are shown in **Figure 1**c, with the panel on the left showing the emission signal from the tandem-NBG top subcells at 1000 nm and the panel on the right showing the tandem-WBG bottom subcells at 680 nm, each under a current of 4 mA. All EL images are shown with auto-scaling of the colorbar values for finer visualization. To compare the performance of all-perovskite tandem subcells with their single junction counterparts, we also performed the EL mapping analyses for the SJ-NBG and SJ-WBG solar cells (**Figure S1**, Supporting Information). The single junction cells exhibit a more homogeneous luminescence distribution compared to the subcells within the tandem solar cells as described later.

The typical external quantum efficiency (EQE) spectra of tandem NBG and WBG subcells (see **Figure S2**, Supporting Information for the reflection spectrum) and the SJ NBG and WBG solar cells are presented in **Figure 1**d. The single junction devices have marginally higher EQEs compared to the top and bottom subcells of the tandem cells. Calculation of the light-generated current density, $J_L$, under the air mass 1.5 global (AM 1.5G) spectrum from these EQE measurements of the SJ devices gives 17.1 mA/cm² for the SJ-WBG and 29.3 mA/cm² for the SJ-NBG cell. For the tandem NBG and WBG subcells, $J_L$ = 14.9 mA/cm² and 15.1 mA/cm², respectively, demonstrating that the two junctions are well-matched in the tandem structure. We have carried out the light *J-V* measurements (AM 1.5G standard reporting condition) on these three different types of solar cells to exhibit the differences between the average device parameters in reverse and forward bias scans, with some typical cell measurements shown in **Figure 1**e.

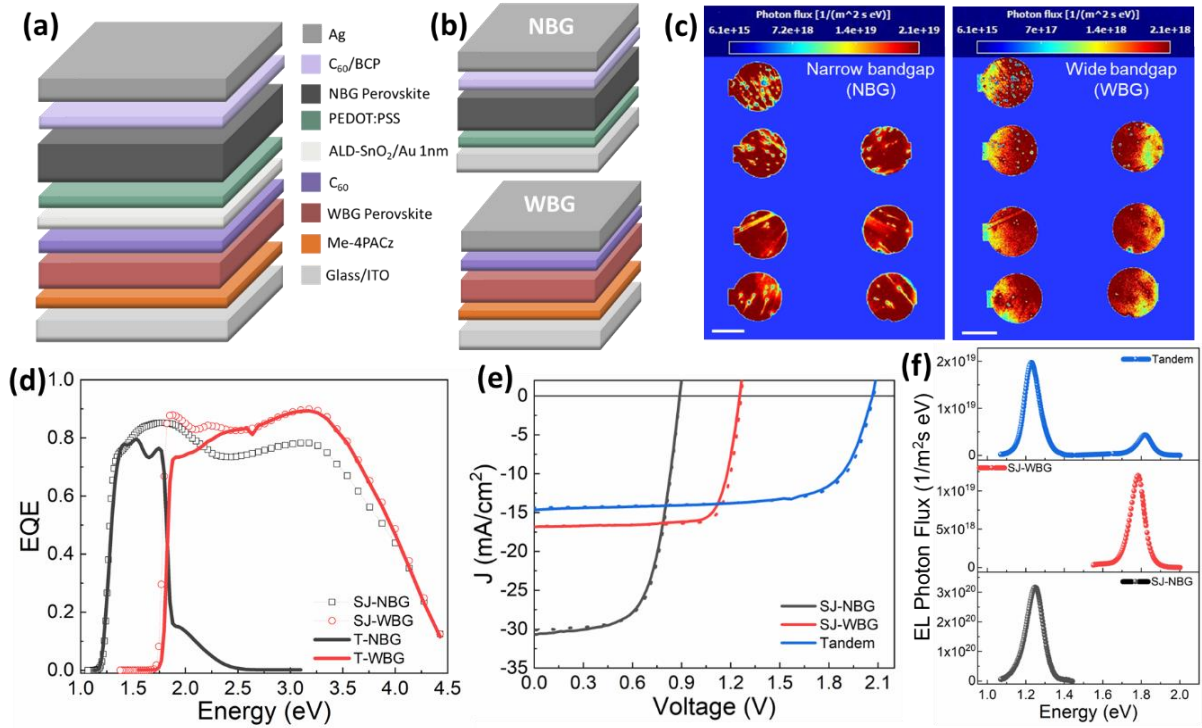

**Figure 1.** (a) Schematic of the two-terminal perovskite tandem solar cell used in this study. (b) Schematics of SJ-NBG and SJ-WBG solar cells. (c) EL images of T-NBG and T-WBG subcells of tandem perovskite cells, showing the nonuniformity in the luminescence map of each subcell and the variability among subcells on the same substrate. Scale bar is 2 mm. (d) EQE spectra of the subcells within tandem solar cells and their comparison with single junction solar cell EQEs. (e) AM 1.5G *J-V* curves of tandem and single junction solar cells in forward (solid line) and reverse (dotted line) directions. (f) Mean EL spectra of tandem and SJ perovskite solar cells obtained from EL images such as those in (c).

The average device parameters -$V_{oc}$, short circuit current density ($J_{sc}$), fill factor (*FF*), and *PCE*- are given in Table 1, and the box plots of the parameters are shown in **Figure S3** (Supporting Information). The average device parameters, particularly for the NBG cells are inferior to our previously reported state-of-the-art devices.[9,10] However, these cells are quite suitable for demonstrating the power of the electroluminescence technique in quantifying nonradiative recombination losses. We note that the $J_{sc}$ of the tandem cells are almost always lower than the current-limiting SJ-WBG device, likely a consequence of some optical losses in the WBG layer closer to the band gap edge due to the absence of metal back electrode as an optical reflector for longer wavelength photons. Another important observation is that the sum of the $V_{oc}$ of the SJs is approximately 100 mV larger than the tandem $V_{oc}$. Therefore, significant nonradiative losses have been introduced into the tandem structure that were otherwise nonexistent in the SJ cells. We will quantify these differences in this work. The origin of these differences could be related to either a high recombination rate or a low tunneling efficiency in the recombination layer in the tandem structure[8] but will ultimately require more investigation.

The *J-V* curves of these three different perovskite devices in the reverse and forward direction show very little hysteresis. Their current-normalized EL spectra are shown in **Figure 1**f. The EL peak positions of the SJ-NBG and the SJ-WBG solar cells are located at approximately 1.25 eV and 1.78 eV, respectively. The EL spectrum of the tandem solar cell exhibits two EL emission peaks at ~1.23 eV (from NBG subcell) and ~1.82 eV (from WBG subcell).

**Table 1.** Average device parameters of single junction NBG (6 devices), WBG (6 devices), and tandem (4 devices) solar cells for the reverse (forward) *J-V* scan.

| Device type | $V_{oc}$ (V) | $J_{sc}$ (mA/cm$^2$) | FF (%) | PCE (%) |
|---|---|---|---|---|
| SJ-NBG | 0.88 (0.87) | 29.44 (29.90) | 63.72 (59.08) | 16.51 (15.48) |
| SJ-WBG | 1.27 (1.26) | 17.31 (17.34) | 81.13 (78.15) | 17.91 (17.08) |
| Tandem | 2.05 (2.05) | 13.91 (14.19) | 69.49 (65.60) | 19.88 (19.07) |

## 2.2. ERE maps

We converted the absolute hyperspectral EL images into EL-ERE maps in order to better understand the spatial variability in emission between the SJ and tandem devices. We then proceeded to calculate the mean and the standard deviation of the ERE values for each image using all pixels at a fixed current density of $J_{inj} = \approx 50$ mA/cm$^2$ (in total, 6 SJ-WBG cells, 6 SJ-NBG cells, and 7 tandem NBG and WBG subcells were analyzed). For each cell type, we determined the ratio of the average standard deviation to the average mean ERE as 16.5 % for SJ-WBG, 38.2 % for T-WBG, 26.1 % for SJ-NBG, and 34.9 % for T-NBG solar cells. We selected an image from each measurement with a ratio closest to these average values and show it below in **Figure 2**a-d. All mapping parameters are given in the Supporting Information (Table S1-S4). In addition, select ERE x and y line scans for each image are shown in **Figure 2**a`-d`. In general, we find a more non-uniform morphology caused by localized defects for the subcells in the tandem structure than their SJ counterparts. These local defects in the subcell layers, likely caused by the inadequate solvent protection of the imperfect atomic layer deposition (ALD) SnO$_2$ interconnecting layer, can contribute to undesirable outcomes as they cause local variability in the nonradiative recombination losses or can act as local shunts.[22] Therefore, a more homogeneous radiative emission is desired. However, the biggest difference observed is the magnitude of the ERE response between the SJ and tandem structures. The average SJ-WBG ERE values are about 12x higher than the T-WBG ERE values and the SJ-NBG EREs are about 18x higher than the T-WBG ERE value. These differences exist even in areas of the images where the ERE is relatively uniform for both types of devices. Therefore, we infer that tandem subcells suffer more significantly from nonradiative losses than their single junction counterparts across their entire device area.

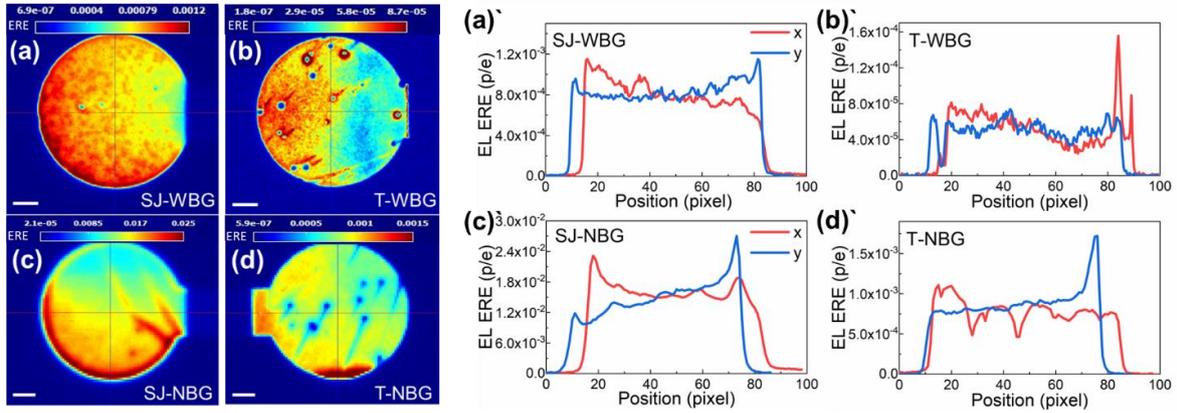

**Figure 2.** ERE mapping of solar cells. ERE maps of (a) a SJ-WBG solar cell, (b) a T-WBG subcell within a tandem cell, (c) a SJ-NBG solar cell, and (d) a T-NBG subcell within a tandem solar cell. Scale bar is 500 µm. (a`, b`, c`, d`) Plots of ERE x- and y- line scans of the 4 ERE device maps as indicated on the maps.

*2.2.1. The mean ERE characteristics*

In previous work,[15] we have shown extensively that EL-ERE can have a substantial current dependence. Therefore, characterizing ERE as a function of the injection current density is both important and potentially insightful. Electroluminescence is considered the reciprocal process to the operation of a solar cell. The ratio of the total external radiative emission current ($qR_{ext}^{tot}$) to the total dark recombination current density ($J_{dark}$) gives the ERE ($\eta_r$) for EL measurements, $\eta_r = qR_{ext}^{tot}/J_{dark}$ where $q$ is the electron charge. In addition to ERE vs. $J$ characterization, we can also plot the calculated ERE as a function of the junction voltage. That way, any negative effects related to series resistance can be accounted for. The electro-optical reciprocity relationship can be utilized to determine the junction voltage, and it is a very powerful relationship when applied to the subcells within a two-terminal tandem cell structure. The reciprocity relationship determines the connection between the spectral external radiative emission rate $R_{ext}$ [E], the external quantum efficiency $\eta_{ext}$ [E], and the junction voltage V, as shown below in Equation 1[18,23]:

$$R_{ext}[E] = \eta_{ext}[E] \frac{2\pi E^2}{h^3 c^2 (exp\left(\frac{E}{kT}\right)-1)} exp\left(\frac{qV}{kT}\right) \quad (1)$$

Here, $h$ is Planck's constant, $c$ is the speed of light in vacuum, $k$ is the Boltzmann constant, $E$ is the photon energy in eV, and $T$ is the device temperature in K. In Equation 1, the absolute EL emission rates, obtained under various injection currents $J_{inj}$, can be used to determine the junction voltage $V$. In these calculations, separately measured EQE curves (**Figure 1**d) have been utilized.

The mean EL-ERE from each of the EL images described in the previous section was plotted as a function of the injection current density and the junction voltage, as shown in **Figure 3** (See **Figure S4** and **S5**, Supporting Information for more detail). For the junction voltage extraction, we fit the mean EL emission rate spectra to Equation 1 and solved for $V$. Error bars in both the x and y coordinates capture the statistical variability across all the recorded EL images. The ERE as a function of $V$ or $J_{inj}$ plots in **Figure 3** reveal that within the tandem structure (**Figure 3**a,d), the NBG subcells are about one order of magnitude more radiatively efficient than the WBG subcells. So the WBG cells suffer more significantly from nonradiative losses, and the $V_{oc}$ deficit (i.e., the voltage loss compared to the radiative voltage limit) in WBG subcells is larger than in the NBG subcells. This ERE difference between the two absorbers, however, is evident in SJ measurements as well, even before incorporation into a tandem stack (**Figure 3**b,c,e,f). An interesting observation here is that while the WBG ERE is about 10× smaller in the subcell compared to the SJ form, this difference grows to 20× for the NBG subcell compared to the NBG SJs. Therefore, the NBG material in the tandem structure degrades more than the WBG material. This finding has major implications for the $V_{oc}$ gap as explained next. Defects causing this extra nonradiative recombination loss in tandem structures likely originate at the failures of the ALD SnO$_2$ interfacial layer (**Figure 1**a), and some are a result of additional localized defect regions as shown in the mapping data (**Figure 2**). We also notice that SJ-NBG ERE values saturate at ≈ 1.4 % at a current density of > 10 mA/cm$^2$ whereas the SJ-WBG EREs continue to increase even at $J_{inj}$ ≈ 60 mA/cm$^2$. A lack of saturation in the ERE plots indicates a more defective material. As the injection current is increased, the Shockley-Read-Hall (SRH) recombination starts to saturate as traps are filled and therefore the radiative recombination component contributes more towards the share of the total recombination losses. When eventually the ERE saturates at higher currents, the cell operates at its maximum external radiative efficiency limit. Internal radiative efficiency can be much higher than the ERE, however, since the ERE also depends on the extraction efficiency from the cell.[24]

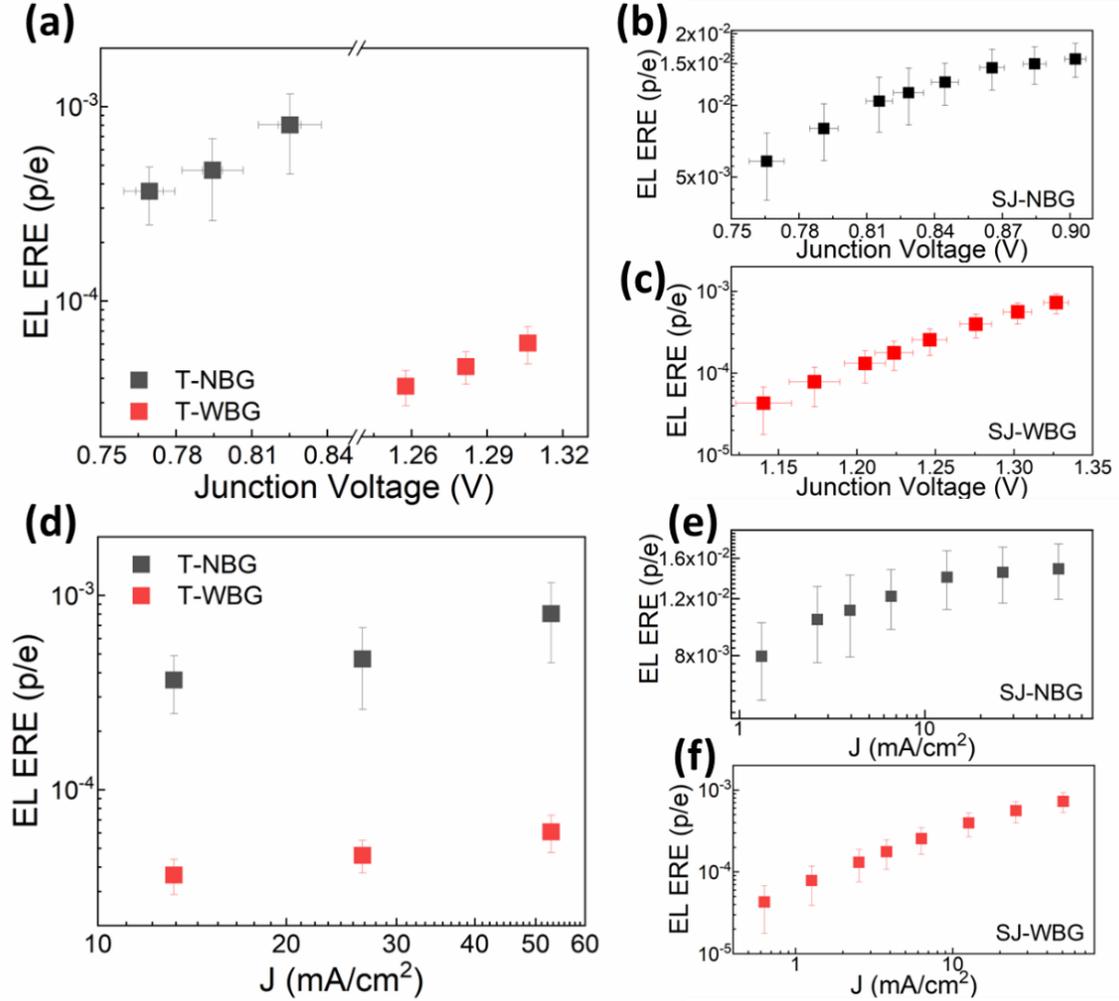

**Figure 3.** Measured ERE vs. V for the (a) tandem solar cells, (b) SJ-NBG solar cells, and (c) SJ-WBG solar cells. Measured ERE vs. $J_{inj}$ for the (d) tandem solar cells, (e) SJ-NBG solar cells, and (f) SJ-WBG solar cells.

## 2.3. Modeling the ERE and *J-V* results

We are interested in using the data shown in **Figure 3** to estimate the voltage losses in each subcell within the tandem 2-terminal structure during the operation of the cell under the standard reporting conditions. Next, we show how we can decompose the blue tandem *J-V* curve in **Figure 1**e into its underlying constituents using the ERE measurements described here.

The dark recombination current in these relatively well-behaved devices can be described by the two-diode model:[18]

$$J_{dark}[V] = J_{01}\left(\exp\left(\frac{qV}{n_1 kT}\right) - 1\right) + J_{02}\left(\exp\left(\frac{qV}{n_2 kT}\right) - 1\right) \quad (2)$$

In this equation, $J_{01}$ and $J_{02}$ are reverse saturation current densities for band-to-band radiative and non-radiative (SRH) recombination mechanisms, respectively, with the ideality factors of $n_1 \cong 1$ and $n_2 \cong 2$. Note that ideality factors can change depending on the transporting layers in solar cells due to interface energetics and the associated charge-carrier capture rates.[25,26] The ERE as a function of the junction voltage, $\eta_r[V]$ can be calculated as the ratio of the radiative component of the dark $J$-$V$ equation multiplied by a photonic limit, $\eta$, to the total of radiative and non-radiative components, as given in Equation 3. The $\eta$ parameter defines the maximum efficiency for extracting light from the cell, and $V$ is the junction voltage calculated by the reciprocity relationship in Equation 1. It is important to note that the junction voltage is different from the voltage measured by a source meter instrument since the measured voltage has an extra voltage contribution due to resistive losses through the contacts and other inactive layers in a cell.

$$\eta_r[V] = \frac{\eta J_{01}\left(\exp\left(\frac{qV}{n_1 kT}\right) - 1\right)}{J_{01}\left(\exp\left(\frac{qV}{n_1 kT}\right) - 1\right) + J_{02}\left(\exp\left(\frac{qV}{n_2 kT}\right) - 1\right)} \quad (3)$$

ERE as a function of $J_{inj}$, $\eta_r[J_{inj}]$, can be calculated from Equation 3 if we solve Equation 2 for $V$ as a function of $J_{dark} = J_{inj}$. This calculation can be done numerically to obtain $V$ as a function of $J_{inj}$, i.e., $V[J_{inj}]$. Therefore, $\eta_r[J_{inj}]$ is given by:

$$\eta_r[J_{inj}] = \frac{\eta J_{01}\left(\exp\left(\frac{qV[J_{inj}]}{n_1 kT}\right) - 1\right)}{J_{01}\left(\exp\left(\frac{qV[J_{inj}]}{n_1 kT}\right) - 1\right) + J_{02}\left(\exp\left(\frac{qV[J_{inj}]}{n_2 kT}\right) - 1\right)} \quad (4)$$

Finally, the $J$-$V$ curves can be modeled using a modified version of Equation 2 that incorporates both the shunt resistance, $R_{sh}$ and the series resistance, $R_s$ into the model:

$$J[V] = -J_L + J_{01}\left(\exp\left(\frac{q(V - R_s A J[V])}{n_1 kT}\right) - 1\right) + J_{02}\left(\exp\left(\frac{q(V - R_s A J[V])}{n_2 kT}\right) - 1\right) + \frac{V - R_s A J[V]}{R_{sh} A} \quad (5)$$

In Equation 5, $A$ is the effective cell area, extracted from EL images; $J_L$ is the light generated current density when $J$-$V$ is measured under light, and $J_L=0$ for a dark $J$-$V$ curve.

We wrote a program to simultaneously fit Equation 3, 4 and 5 to all four plots of ERE vs $V$, ERE vs $J_{inj}$, and $J$-$V$ (light and dark) curves. Such an approach allows for the most accurate and robust extraction of the fit parameters, since four of the fit parameters, namely $J_{01}$, $J_{02}$, $n_1$ and $n_2$, must satisfy all four data plots. The parameter $\eta$ is only valid for the two ERE plots, and $R_s$ and $R_{sh}$ are only relevant for the $J$-$V$ curves.

*2.3.1. Single junction modeling results*

We first validated this method on the data obtained from the SJ NBG and WBG solar cells, with the results shown in **Figure 4**a-d for one such pair of devices, and the corresponding model parameters given in Table 2. Good agreement was observed between the data and the model for these single junction cells. The dependence of the ERE on current or voltage is somewhat related to the $J_{02}/J_{01}$ ratio. For the WBG cell, this ratio is $\approx 10^{11}$ whereas this ratio is reduced to $\approx 10^6$ for the NBG cell. A smaller ratio corresponds to a weaker dependence of ERE on $J_{inj}$ and an indication that the cell suffers less from nonradiative defects than a higher ratio case. The higher EREs observed in NBG cells, compared to WBG cells, can be attributed to high quality NBG films with fewer bulk and interfacial defects for these devices. Furthermore, the $J_{01}$ and $J_{02}$ parameters are critical in determining the correct $V_{oc}$ for light $J$-$V$ curves, and the excellent fit to all 4 data sets validates the accuracy of the ERE measurements.

Comparing $R_s$ and $R_{sh}$ values between the two cells, we find the NBG cells suffer from a higher series resistance, which also seems to be the major reason for their lower *FF* (see Table 1). These series resistances do not change between light and dark $J$-$V$s, but the shunt resistances do seem to be light-intensity dependent and decrease substantially for light $J$-$V$s, a common behavior that has been documented in solar cells previously.[27,28] It is important to note here that even though the *PCE* of our NBG cells are slightly lower than the WBG cells, this difference is entirely attributable to the lower *FF* of the NBG cells. If the *FF* of our NBG cells were the same as WBG cells, i.e., $\approx 81$ % vs 64 %, then the *PCE* of the NBG cells would be 27 % higher (81/64), or about 22.7 %. Also, we draw attention to the voltage deficit, $V_{def}$, for the two cells. $V_{def}$ is a voltage penalty term that reduces the observed $V_{oc}$ from its maximum radiative limit, and is given by:

$$V_{\text{def}} = \frac{kT}{q}\ln(\eta_r) \tag{6}$$

Notice that Equation 6 produces a negative number since $\eta_r \leq 1$. Given an $\eta_r = 0.014$ for the NBG cells at $J \approx 30$ mA/cm$^2$, $V_{def,\,NBG} = 110$ mV, compared to $V_{def,\,WBG} = 192$ mV at $J \approx 15$ mA/cm$^2$. Therefore, the smaller voltage deficit for the NBG cells combined with better fill factors have allowed these cells to produce champion PCEs of $\approx 23$ % previously.[9]

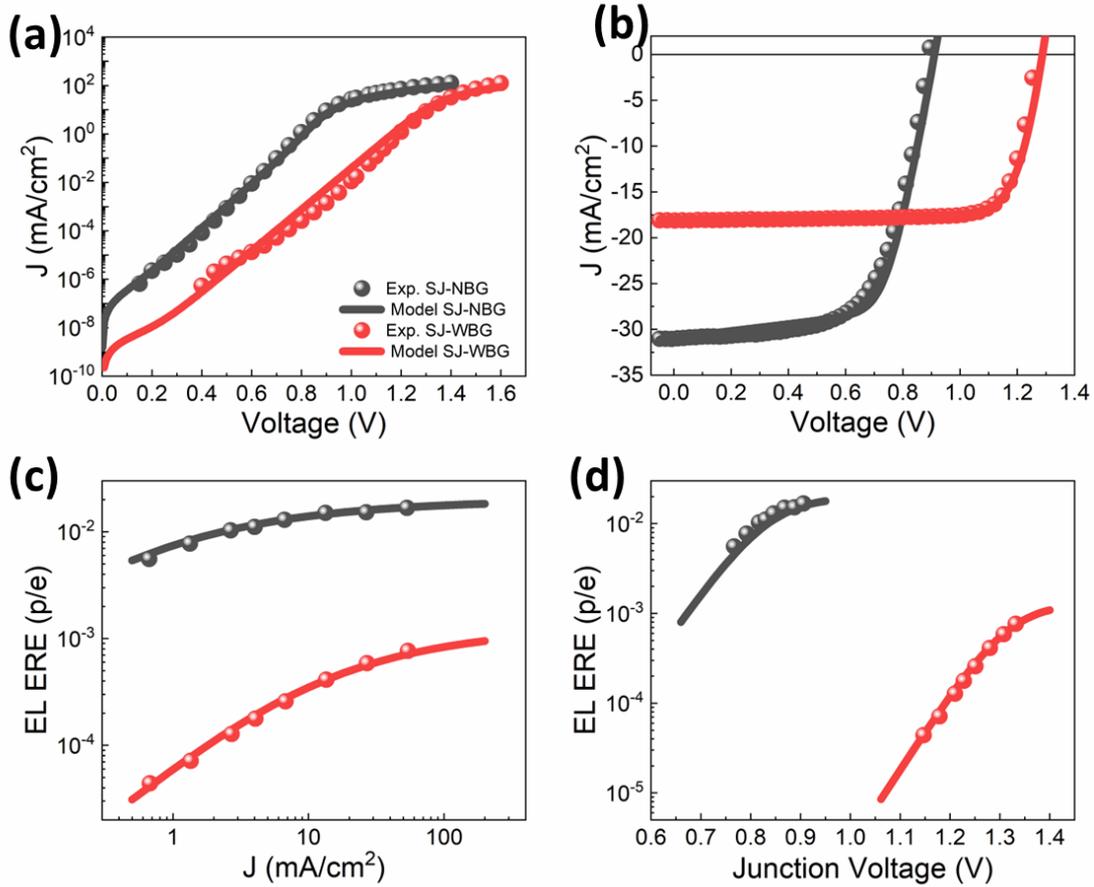

**Figure 4.** NBG and WBG single junction solar cells. (a) Dark and (b) light *J-V* plots with the 2-diode model. (c) Measured ERE vs. $J_{inj}$ and (d) Measured ERE vs. V with the ERE model.

**Table 2.** Two-diode model fit parameters for the NBG and WBG SJ solar cells.

| Device type | η | $J_L$ (mA/cm$^2$) | $J_{01}$ (mA/cm$^2$) | $J_{02}$ (mA/cm$^2$) | $n_1$ | $n_2$ | $R_s$ (Ω.cm$^2$) | $R_{sh}$ (Ω.cm$^2$) |
|---|---|---|---|---|---|---|---|---|
| SJ-NBG dark | 0.02 | 0 | 7.9x10$^{-15}$ | 3.2x10$^{-8}$ | 1 | 1.873 | 4.72 | 7.5x10$^8$ |
| SJ-WBG dark | 0.0013 | 0 | 6.3x10$^{-22}$ | 1.1x10$^{-10}$ | 1 | 2 | 2.96 | 3.7x10$^{10}$ |
| SJ-NBG light | 0.02 | 31.5 | 7.9x10$^{-15}$ | 3.2x10$^{-8}$ | 1 | 1.873 | 4.72 | 262 |
| SJ-WBG light | 0.0013 | 18.0 | 6.3x10$^{-22}$ | 1.1x10$^{-10}$ | 1 | 2 | 2.96 | 2.3 x10$^3$ |

Defects such as pinholes provide a contribution to the decrease in shunt resistance or an increase in ideality factor.[22]

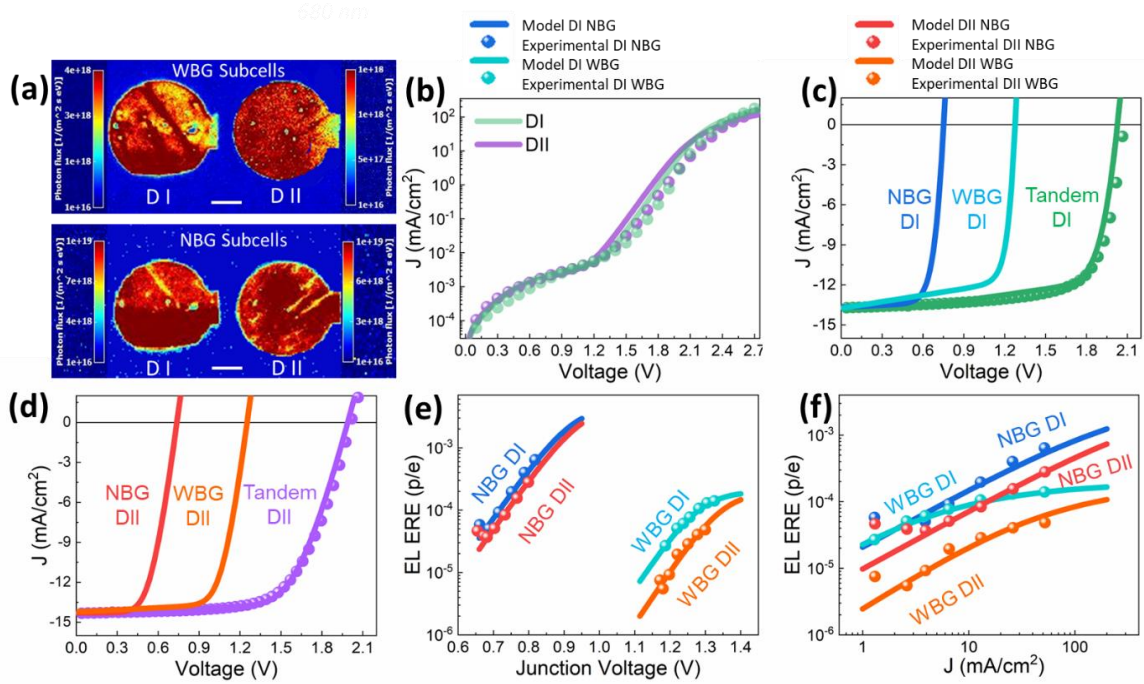

**Figure 5.** Comparison of tandem solar cells. (a) Absolute hyperspectral images of measured subcells of tandem solar cells. Bar size is 1 mm. (b) Dark *J-V* plots with the 2-diode model of DI and DII tandem solar cells. (c) Light *J-V* plots with the 2-diode model of DI tandem device and its subcells. (d) Light *J-V* plots with the 2-diode model of DII tandem device and its subcells. (e) Measured ERE as a function of V and (f) measured ERE as a function of $J_{inj}$ with the ERE model of tandem solar cells.

**Table 3.** Device parameters of tandem solar cells in reverse (forward) direction.

| Device name | $V_{oc}$ (V) | $J_{sc}$ (mA/cm$^2$) | FF (%) | PCE (%) |
|---|---|---|---|---|
| DI | 2.08 (2.04) | 13.73 (14.05) | 71.64 (66.19) | 20.43 (18.99) |
| DII | 2.01 (2.02) | 14.33 (14.43) | 65.48 (63.81) | 18.88 (18.65) |

*2.3.2. Tandem cells modeling results*

We next turn to ERE and *J-V* analysis of the tandem structures using the two-diode model. For this analysis, we picked two different devices, DI and DII, with slightly different *J-V* curve parameters as shown in Table 3. In particular, DI produced a $V_{oc}$ that is 70 mV higher than DII, and with a slightly higher FF, it produces a PCE of 20.43 %. **Figure 5**a shows the EL images; experimental and modeled dark *J-V* and light *J-V* curves are shown in **Figure 5**b-d, and the measured and modeled subcell EREs are shown in **Figure 5**e,f. Equation 3-5 were simultaneously fit to all of the shown data and the individual subcell parameters are listed in Table 4. One important difference between the modeling of single junction cells (**Figure 4**) and tandem cells (**Figure 5**) is that experimental subcell *J-V* data is not available for tandems since they are two-terminal devices. However, individual subcell EREs are measured and therefore are utilized in the modeling process to constrain the fit parameters and arrive at modeled subcell *J-V* curves as shown by solid lines in **Figure 5**c,d. When these modeled WBG and NBG subcell *J-V*s are added in series, they reproduce the composite experimental *J-V* curves reasonably well, both under light and in the dark. Small differences between the model and the data can be attributed to the uncertainty in the ERE measurements and degradation effects (dark and light *J-V*s and voltage-dependent ERE measurements using a two-camera system are carried out sequentially over a span of several days and can induce bias stress on the devices).

We can now compare the modeled subcell *J-V* curve parameters with single junction parameters to understand how extra charge carrier recombination in a tandem structure affects the performance of each subcell. Here, we focus primarily on the $V_{oc}$ losses in each subcell and ignore the $J_{sc}$ and *FF* differences because those parameters are not as directly related to nonradiative recombination losses. $J_{sc}$ of a tandem cell is generally determined by the current-limiting subcell, i.e., the WBG cell, and any optical losses through the WBG layers as we discussed earlier. The *FF* is impacted by series and shunt resistances. Previous work has shown that obtaining an 80 % FF in these tandem structures is quite achievable.[9] So the $V_{oc}$ gap is a more important issue to investigate. Comparing the ratios $J_{02}/J_{01}$ between subcell and SJ WBG and NBG devices, we find that the subcell WBG ratios are still $\approx 10^{11}$, similar to previous SJ WBG findings. However, these ratios increase to $\approx 3 \times 10^7$ for the NBG subcells compared to $10^6$ for SJ NBGs. As discussed earlier, a higher ratio corresponds to a deterioration in the radiative channels and more nonradiative recombination losses for the NBG subcell. We also observe further reductions in the $R_{sh}$ values for the subcells, but these reductions are not large enough to impact the $V_{oc}$ or the $J_{sc}$ in any meaningful way. Finally, with the $R_s$ parameter, it is not possible to attribute distinct values to each subcell because the subcell *J-V*s are added in series and only the sum of the two $R_s$ values needs to fit the tandem *J-V* curve. Therefore, we attributed an equal $R_s$ value to both subcells.

Comparing the $V_{oc}$ values obtained through the modeling of the DI and DII tandem subcells with the average NBG and WBG SJ cells (see Table 5), we find that the $\Delta V_{oc}$ between the WBG subcells and the WBG SJ cells are generally very small and under 15 mV, whereas $\Delta V_{oc}$ between the NBG subcells and the NBG SJ cells are in the range of 100 mV to 150 mV (see column 2). We can also calculate the $\Delta V_{oc}$ term based on our ERE measurements and an estimate of the band gap energy, $E_g$. In general, $V_{oc}$ is given by:

$$V_{oc} = \frac{E_g}{q} - \Sigma V_{loss} + \frac{kT}{q}\ln(\eta_r) \tag{7}$$

Here, $\frac{kT}{q}\ln(\eta_r)$ defines entropic radiative loss, and $\Sigma V_{loss}$ is the sum of the other entropic losses (the solid angle of the sun, photon cooling, and density of state modifications) that the cell incurs,[29] the biggest of which is related to the mismatch between the incident solar solid angle and the cell's emission solid angle, and the last term is the voltage deficit term. The $\Delta V_{oc}$ between the SJ $V_{oc}$ measurements and the subcell model-derived values is therefore given by:

$$\Delta V_{oc} = \frac{E_g^{SJ}}{q} + \frac{kT}{q}\ln(\eta_r^{SJ}) - \frac{E_g^{subcell}}{q} - \frac{kT}{q}\ln(\eta_r^{subcell}) \tag{8}$$

We have included $E_g$ values in this calculation because, in practice, we have observed that $E_g$ values can show small variations between nominally similar cells and also between the subcell and single junction WBG and NBG devices. For $E_g$ estimates, we used the peak luminescence emission energies from the EL measurements as shown in **Figure 1**f and **S6** (Supporting Information). These estimates along with the ERE values at a current density of ≈ 13 mA/cm² and the calculated $\Delta V_{oc}$ based on Equation 8 are also shown in Table 5. The observed and the calculated $\Delta V_{oc}$ values generally agree well for all 4 comparisons. This agreement further validates the modeling approach that we have taken to derive the subcell *J-V* curves for the tandem solar cells.

**Table 4.** Two-diode model fit parameters of NBG and WBG subcells of tandem solar cells.

| Device type | η | $J_L$ (mA/cm$^2$) | $J_{01}$ (mA/cm$^2$) | $J_{02}$ (mA/cm$^2$) | $n_1$ | $n_2$ | $R_s$ (Ω.cm$^2$) | $R_{sh}$ (Ω.cm$^2$) |
|---|---|---|---|---|---|---|---|---|
| DI NBG-dark | 0.0049 | 0 | 7.24x10$^{-14}$ | 1.78x10$^{-6}$ | 1 | 1.873 | 1.54 | 1.2 x 10$^6$ |
| DI WBG-dark | 0.0002 | 0 | 8.71x10$^{-22}$ | 7.76x10$^{-11}$ | 1 | 2 | 1.54 | 1.9 x 10$^5$ |
| DI NBG-light | 0.0049 | 13.7 | 7.24x10$^{-14}$ | 1.78x10$^{-6}$ | 1 | 1.873 | 1.93 | 1.9 x 10$^3$ |
| DI WBG-light | 0.0002 | 13.8 | 8.71x10$^{-22}$ | 7.76x10$^{-11}$ | 1 | 2 | 1.93 | 6.1x 10$^2$ |
| DII NBG-dark | 0.0049 | 0 | 7.08x10$^{-14}$ | 2.63x10$^{-6}$ | 1 | 1.873 | 2.44 | 9.6 x 10$^5$ |
| DII WBG-dark | 0.0002 | 0 | 7.24x10$^{-22}$ | 2.40x10$^{-10}$ | 1 | 2 | 2.44 | 2.4 x 10$^5$ |
| DII NBG-light | 0.0049 | 14.3 | 7.08x10$^{-14}$ | 2.63x10$^{-6}$ | 1 | 1.873 | 10.01 | 1.2 x 10$^5$ |
| DII WBG-light | 0.0002 | 14.3 | 7.24x10$^{-22}$ | 2.40x10$^{-10}$ | 1 | 2 | 10.01 | 1.9 x 10$^3$ |

**Table 5.** Comparison of the observed and derived $V_{oc}$ values between SJ and tandem solar cells.

| Device name | $V_{oc}$ (V) | $\Delta V_{oc}$ (mV) observed | $E_g$ (eV) | ERE | $\Delta V_{oc}$ (mV) calculated |
|---|---|---|---|---|---|
| WBG SJ cell | 1.268 | | 1.781 | $4 \times 10^{-4}$ | |
| DI-WBG subcell | 1.278 | -10 | 1.822 | $1.1 \times 10^{-4}$ | -7 |
| DII-WBG subcell | 1.253 | 15 | 1.822 | $2.9 \times 10^{-5}$ | 27 |
| NBG SJ cell | 0.878 | | 1.249 | 0.014 | |
| DI-NBG subcell | 0.750 | 128 | 1.230 | $1.9 \times 10^{-4}$ | 130 |
| DII-NBG subcell | 0.738 | 140 | 1.230 | $8.4 \times 10^{-5}$ | 152 |

## 3. Discussion

Our measurements and modeling results indicate that although the individual NBG solar cells are more radiatively efficient than their WBG counterparts, their initial radiative efficiency advantage can suffer significantly when incorporated in a tandem cell structure. For example, the WBG subcell in the DI tandem cell shrinks the WBG voltage gap by 10 mV ($\Delta V_{oc}$ = -10 mV), mostly due to its slightly larger $E_g$ in the subcell form; however, the NBG subcell adds 128 mV to the voltage gap, most of which is a result of its substantially lower ERE (0.014 in the SJ form vs $1.9 \times 10^{-4}$ in the subcell form). Therefore, a large amount of the voltage gap that exists between the tandem cells and the sum of the individual SJ cells is caused by a substantial amount of nonradiative losses that are introduced in the NBG subcell. A similar trend is observed for the DII device with even larger losses in the NBG subcell. Recent work has shown that the voltage gap can be reduced to ≈ 0 V in very high-quality champion all-perovskite solar cells[10] but in most typical devices, this voltage gap exists and can be large.[9,30–32]

The primary structural distinction between single-junction solar cells and the subcells within a tandem solar cell lies in the presence of the recombination layer. In the tandem configuration, the NBG subcell is deposited on the top of the WBG layers, with the two separated by a recombination layer (ALD $SnO_2$-Au) designed to facilitate efficient charge balance across the device.[8] However, defects within the recombination layer can impair the uniform deposition and crystallization of the overlying NBG subcell.[33] Such defects may promote the formation of interfacial imperfections at the junction between the subcells, thereby degrading overall device performance and contributing to additional voltage losses. On the other hand, these defects in the recombination layer induce Shockley-Read-Hall recombination, an effect that can, under certain conditions, be beneficial in tandem cells by enhancing non-radiative recombination. However, the impact of defect-assisted recombination on the crucial minority charge carriers is required to be minimized since minority carrier recombination is entirely undesirable and must be avoided, as discussed in previous studies.[34,35] Such recombination processes result in voltage losses, highlighting the importance of effective passivation strategies. In such a situation, the extra recombination losses are mostly interfacial and not bulk related. However, if impurities or native defects migrate into the bulk, then bulk recombination losses can also be a significant source of the nonradiative losses. The NBG material starts as a higher radiative efficiency material and can therefore be more prone to penetration by impurities or other defects and can degrade more significantly than the WBG material.

## 4. Conclusion

Absolute EL-based ERE measurements were applied to subcells within all-perovskite tandem solar cells to better understand nonradiative recombination losses in these devices. We observed that radiative efficiencies of both the WBG and NBG subcells are substantially lower compared to single junction WBG and NBG solar cells. However, even though the NBG subcell remains the more efficient one, it degrades more substantially in the tandem structure compared to the WBG cell. Use of the two-diode model on both the ERE and *J-V* data allowed us to calculate individual subcell *J-V* curves under the standard reporting conditions and compare their characteristics, particularly the open circuit voltages, with the single junction cells. Surprisingly, we find that the NBG subcell contributes the most towards the voltage gap between a tandem cell and the sum of its constituents in single junction form even though it is the more efficient subcell. Our results have significant implications for the processing and optimization of all-perovskite tandem solar cells as they reveal that more efforts have to be put towards interface passivation and defect mitigation strategies targeting the NBG instead of the WBG subcell.

The EL hyperspectral imaging and $J$ or $V$ dependent-ERE calculations employed in this work are broadly applicable beyond the tandem solar cells studied here. This technique can also be readily implemented across different perovskite compositions, device architectures, and in other photovoltaic technologies. Their stability under varying environmental conditions and scalability from single cells to modules makes them particularly powerful for both laboratory diagnostics and industrial quality control. Consequently, the methodology studied here provides a generalizable framework for evaluating external radiative efficiency and identifying radiative / non radiative recombination pathways in a wide range of optoelectronic devices.

## 5. Experimental Section

*Fabrication of single junction and all perovskite tandem solar cells:* All perovskite tandem solar cells were fabricated following our previous work.[9] WBG perovskite precursor solution was prepared by dissolving 0.96 mol/L formamidinium iodide (FAI) 0.24 mol/L cesium iodide (CsI), 0.48 mol/L lead iodide ($PbI_2$), and 0.72 mol/L lead bromide ($PbBr_2$) in 1 ml N,N-dimethylformamide (DMF) / dimethyl sulfoxide (DMSO) solvent (volume ratio = 4:1). NBG perovskite precursor solution was prepared by dissolving 0.3 mol/L CsI, 1.9 mol/L FAI, 0.45 mol/L methylammonium Iodide (MAI), 0.9 mol/L $PbI_2$, 0.9 mol/L tin iodide ($SnI_2$), and 0.09 mol/L tin fluoride ($SnF_2$) in 1 ml DMF/DMSO solvent (volume ratio = 3:1). To fabricate tandem devices, nickel oxide ($NiO_x$) nanoparticles were first spin-coated on cleaned indium tin oxide (ITO) substrates at 4000 rpm (1.00 rpm = 0.105 rad/s) for 30 s. The substrates were transferred into a glovebox to coat a [4-(3,6-dimethyl-9*H*-carbazol-9-yl)butyl]phosphonic acid (Me-4PACz) film at 3000 rpm for 25 s and annealed at 100 °C for 10 min. A WBG perovskite film was then deposited on the substrate at a speed of 4000 rpm for 30 s, with 150 μL anisole as antisolvent dropping on the film at 22 s, followed by annealing at 110 °C for 15 min. Then propane-1,3-diammonium iodide ($PDAI_2$) was dynamically coated on the WBG perovskite as surface passivation, followed by annealing at 100 °C for 5 min. After that, the substrates were transferred into an evaporation chamber to deposit a 30 nm $C_{60}$. A polyethylenimine ethoxylated (PEIE) solution was spin-coated on top of $C_{60}$ at 5000 rpm for 30 s. Then, a 25 nm $SnO_2$ film was deposited by atomic layer deposition, and 1 nm Au was deposited by thermal evaporation.

Diluted PEDOT:PSS solution (1:2 in isopropanol) was spin coated at 3000 rpm for 10 s in the clean room (controlled humidity below 30%) and further dried at 120 °C for 8 min. The samples were transferred back to a nitrogen-filled glovebox. An NBG perovskite film was coated at 800 rpm for 5 s and 3500 rpm for 40 s, with 400 μL of chlorobenzene as the antisolvent dropping on the film at 20 s before the end of spin procedures. The NBG film was passivated with ethane-1,2-diammonium (EDA) treatment. Lastly, $C_{60}$, bathocuproine (BCP), and Ag evaporation were used to complete the tandem devices. Before measurements, the devices were packaged with UV epoxy in a glovebox.

*Electroluminescence Measurements:* EL measurements of the perovskite solar cells were performed using the Grand EOS hyperspectral wide-field imaging system by Photon ETC under different applied currents using a Source Meter. This system is comprised two camera systems, a sCMOS camera for measurements in the visible and near infrared (VIS-NIR) region between 400 nm to 1000 nm and a cooled InGaAs camera for spectral measurements in the short-wave infrared (SWIR) region between 850 nm to 1600 nm. The spectral resolution is about 2 nm for the VIS-NIR, and 4 nm for the SWIR ranges. All images were taken by a 20 mm x 20 mm wide field objective lens. The calibration of this system to obtain absolute photon flux rates is discussed elsewhere.[19] The relative uncertainties of the EL flux rates presented here are about ± 15 %. The effective mean area of the perovskite solar cells calculated from their EL images is $7.67 \times 10^{-2}$ cm.

*EQE Measurements:* EQE measurements were carried out in the range of 280 nm to 1160 nm for SJ-NBG, 280 nm to 900 nm for SJ-WBG, 400 nm to 1100 nm for tandem NBG subcell, and 280-800 nm for tandem WBG subcell solar cells using a monochromator-based differential spectral responsivity system. For the tandem devices, appropriate light bias LEDs were used to measure the EQE response of each subcell. The details of the system operation have been reported in our previous work.[36]

*Current vs voltage measurements:* The dark current vs voltage (*I-V*) measurements were performed near room temperature (T = 23 °C) using the steady state I-V approach[37] with Keithley model 2601 source-measure unit. In this method, the cell is held under a given voltage until a relative stability level for the current is achieved before the current is recorded. The voltage sweep values are predefined in a fixed array and the program goes through the list one voltage value at a time. Under each voltage, the cell current is monitored at a rate of about 10 samples/s and the average of every 50 samples is continuously compared against the prior 50-sample average. Once the rate of change for the current comparisons falls below a user-defined value (called the stability value), a current is recorded for the given voltage. We usually require this stability value to be under 2 %. Forward and backward light I-V sweeps for the single junction cells were performed under a xenon solar simulator against an internally calibrated, silicon reference solar cell under the standard reporting conditions (i.e., 1000 W/m$^2$). A spectral mismatch correction parameter was calculated and applied to each type of solar cell I-V measurement. The light I-V sweeps are typically conducted at a rate of 17 ms per point and no device preconditioning was performed. For the tandem cell *I-V*s, the measurements were performed under an adjustable light emitting diode (LED) based solar simulator using two separate NIST-calibrated reference solar cells (A gallium arsenide reference cell for the NBG measurements and an indium gallium phosphide reference cel for the WBG measurements) for setting the effective irradiance at the SRC. The *I-V* results presented here have uncertainties in the range of 1 % to 2 %.

*2-diode model fitting:* Data fitting for each curve was performed using the classical chi-square fitting algorithm proposed by Pearson within the framework of the two-diode model.[38] The resulting standard deviation for the ERE curves lies in the range of 10-50%, while for the *J-V* curves it is higher, on the order of 100 %.

**Acknowledgements**

H.Y-C. acknowledges the generous support of the Professional Research Experience Program (PREP) under the award number PREP0002351. The work at the University of Toledo was supported by the U.S. Department of Energy's Office of Energy Efficiency and Renewable Energy (EERE) under Hydrogen and Fuel Cell Technologies Office Award Number DE-EE0010740, and by the U.S. Air Force Research Laboratory under agreement number FA9453-19-C-1002. S.M.T. acknowledges funding from the National Science Foundation (ECCS-1846239) and support from the Ralph S. O'Connor Sustainable Energy Institute (ROSEI) at Johns Hopkins University. The views expressed in the article do not necessarily represent the views of the DOE or the U.S. Government. The U.S. Government retains and the publisher, by accepting the article for publication, acknowledges that the U.S. Government retains a nonexclusive, paid-up, irrevocable, worldwide license to publish or reproduce the published form of this work, or allow others to do so, for U.S. Government purposes. NIST disclaimer: Certain equipment, instruments, software, or materials are identified in this paper in order to specify the experimental procedure adequately. Such identification is not intended to imply recommendation or endorsement of any product or service by NIST, nor is it intended to imply that the materials or equipment identified are necessarily the best available for the purpose.